\documentclass[12pt]{iopart}
\usepackage{epsfig}
\def\be{\begin{equation}}
\def\ee{\end{equation}}
\def\bea{\begin{eqnarray}}
\def\eea{\end{eqnarray}}

\begin{document}
\begin{flushright}
IFT-2001/36 \\ [1.5ex]
{ \large \bf hep-ph/0110250 } \\
\end{flushright}
 \title[Photoproduction of isolated photons at HERA]
{Photoproduction of isolated photons at HERA in NLO QCD}

\author{Maria Krawczyk \dag\   
\footnote[3]{Invited talk at RINGBERG Workshop: New Trends in HERA Physics 2001\\ and
at Workshop on Small x Physics 2001, Cracow, June 2001}
}
\address{\dag\ Institute of Theoretical Physics, Warsaw University,
ul. Hoza 69,  00-681 Warsaw, Poland}
\begin{abstract}
{ The NLO QCD calculation for the photoproduction of the isolated photon with 
a large $p_T$ at the HERA $ep$ collider is presented. The single resolved 
photon contribution and the QCD corrections 
of order $\alpha_s$ to the Born term are consistently included. The sizeable
NNLO contributions, the box and the double resolved photon subprocesses, are 
  taken into account in addition. The importance 
of the isolation cut, as well as the influence of other experimental cuts on 
the $p_T$ and $\eta_{\gamma}$ 
distributions are discussed in detail. 
Results are compared with experimental data and with the  
different NLO calculations.}
\end{abstract}
\maketitle
\section{Introduction}
The production in the $ep$ collision
of the prompt photon with large transverse momentum $p_T$ 
 is considered.
Such reaction is dominated by events with 
almost real photons mediating the $ep$ interaction, 
$Q^2\approx 0$, so in practice we deal with the photoproduction of the 
prompt photon (called also  
Deep Inelastic Compton (DIC) scattering). 
The photon emitted by the electron
may interact with the proton partons directly
or as a resolved one. Analogously, 
the observed final photon
may  arise directly from hard partonic subprocesses 
or from fragmentation processes, 
where a quark or a gluon decays into $\gamma$.

The importance of the DIC process in the $ep$ 
collision for testing the Parton Model and then the Quantum Chromodynamics 
was studied previously by many
authors~\cite{Bjorken:1969ja}-\cite{Gordon:1994sm}.
Measurements were performed at the HERA $ep$ collider by the ZEUS 
group~\cite{Breitweg:1997pa}-\cite{Breitweg:2000su}, and \cite{Chekanov:2001aq},
also the H1 Collaboration has presented preliminary
results~\cite{h1}. In these experiments  events with isolated photons
were included in the analysis, 
i.e. with a restriction imposed on the hadronic energy 
detected close to the photon. 
The corresponding cross sections for the photoproduction 
of an isolated photon and of an isolated photon plus jet were calculated
in QCD in next-to-leading order 
(NLO)~\cite{Gordon:1995km}-
\cite{Fontannaz:2001nq}.
There  exists analogous calculation for the large-$Q^2$ 
case (DIS events)~\cite{Kramer:1998nb}.

In this talk  the results of the NLO QCD calculation for the 
DIC process with an isolated photon at the HERA $ep$ collider
are presented \cite{Krawczyk:1998it,zk}. We consider the parton 
distributions in the photon and parton fragmentation into the photon
as quantities of order $\alpha_{em}$.
We emphasize the importance of the box diagram $\gamma g\rightarrow \gamma g$,
being the higher order process, in description of the data.
Our approach differs from the NLO
approach~\cite{Gordon:1995km}-\cite{Gordon:1998yt,
Fontannaz:2001ek,Fontannaz:2001nq} 
by set of subprocesses included in the analysis. 
The comparison of our predictions (KZ)~\cite{zk} with the results obtained by 
L.E. Gordon (LG)~\cite{Gordon:1998yt} and M. Fontannaz et al. (FGH)
~\cite{Fontannaz:2001ek} and with data   measured by
 ZEUS group ~\cite{Breitweg:2000su} is presented.

\section{The NLO calculation for $\gamma p\rightarrow\gamma X$ 
--  general discussion}\label{sec:dic}
\begin{figure}
\begin{center}
\epsfysize=3cm
\epsfbox{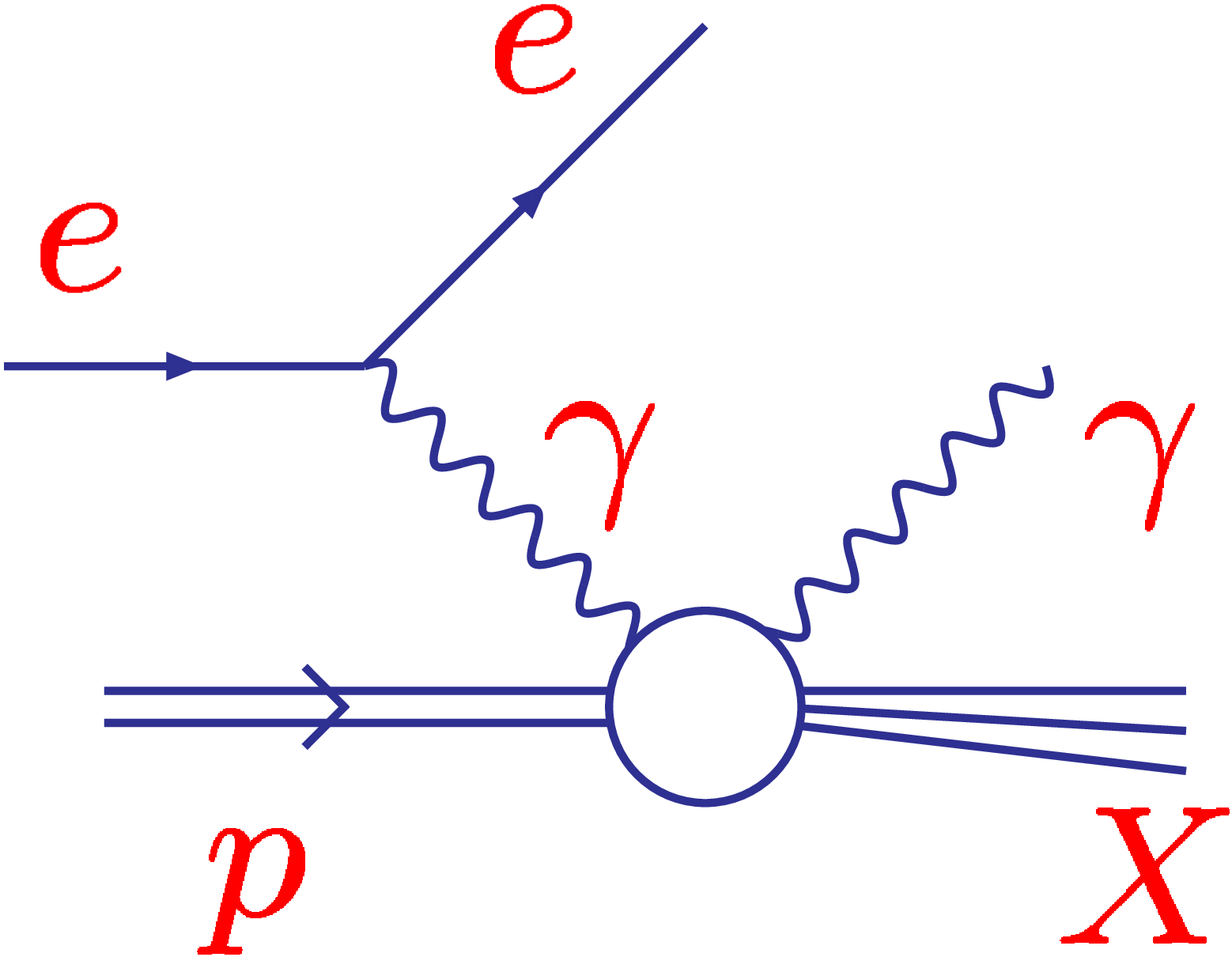}
\hskip 3cm
\epsfysize=3cm
\epsfbox{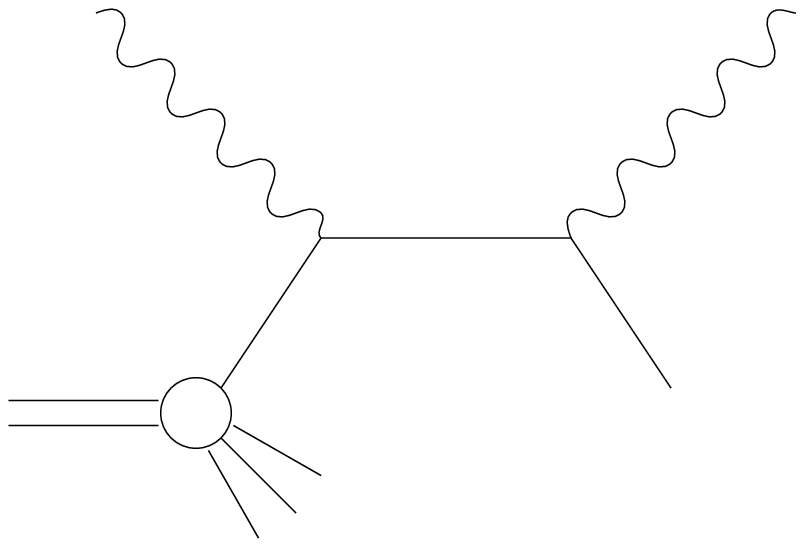}
\end{center}
\caption{\label{fig:dicborn}The Deep Inelastic Compton process (left)
and   Born process (right).}
\end{figure}

Different approaches to the    
NLO calculations of cross sections for hadronic processes involving resolved 
photons exist in literature, see discussion in
~\cite{Krawczyk:1990nq,Krawczyk:1998it} 
and ~\cite{chyla}. Here 
we discuss how  the NLO QCD calculations, based on the DGLAP approach,
are being performed for  the DIC cross section  (fig.~\ref{fig:dicborn},left), 
\be
\gamma p \rightarrow \gamma X,
\label{eq:1}
\ee
where the final photon is produced with large transverse momentum, 
$p_T\gg\Lambda_{QCD}$.

\begin{figure}
\begin{center}
\hspace*{-1cm}
\epsfysize=3.5cm
\epsfbox{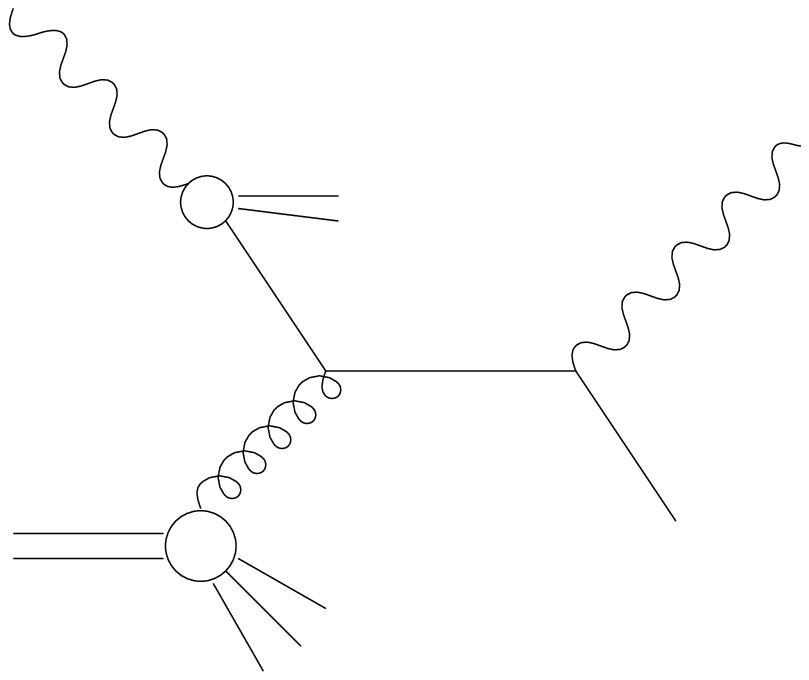}
\epsfysize=3.5cm
\hskip -3.5cm
\epsfbox{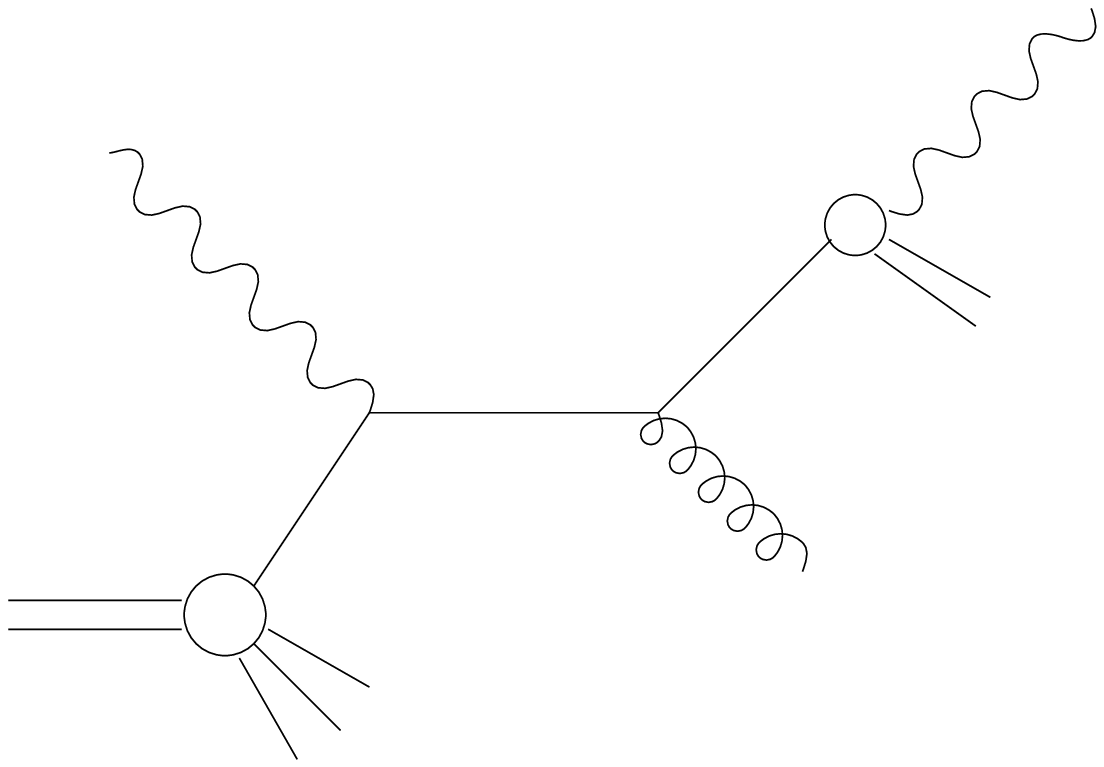}
\epsfysize=3.5cm
\hskip 4cm
\epsfbox{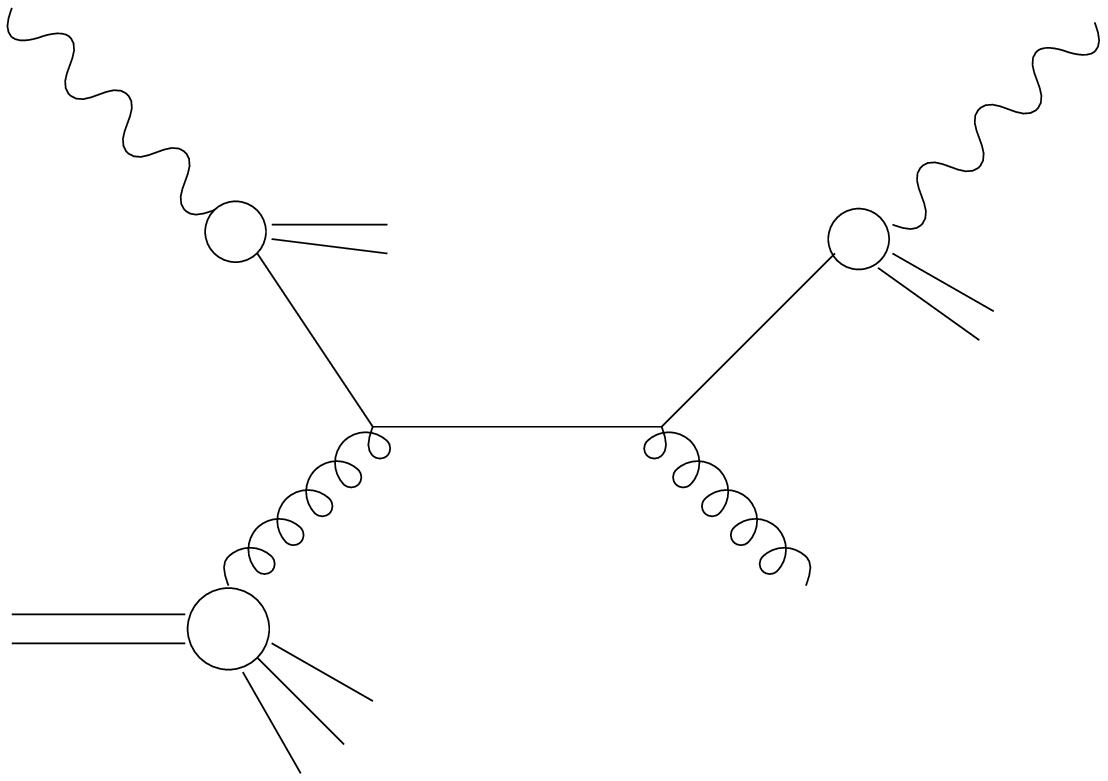}
\end{center}
\caption{\label{fig:12res}Examples of single resolved $\gamma$ processes:
 the resolved initial photon (left) and  the resolved final photon (center).
An example of a double resolved photon process (right).}
\end{figure}

\begin{figure}
\begin{center}
\epsfysize=3.5cm
\epsfbox{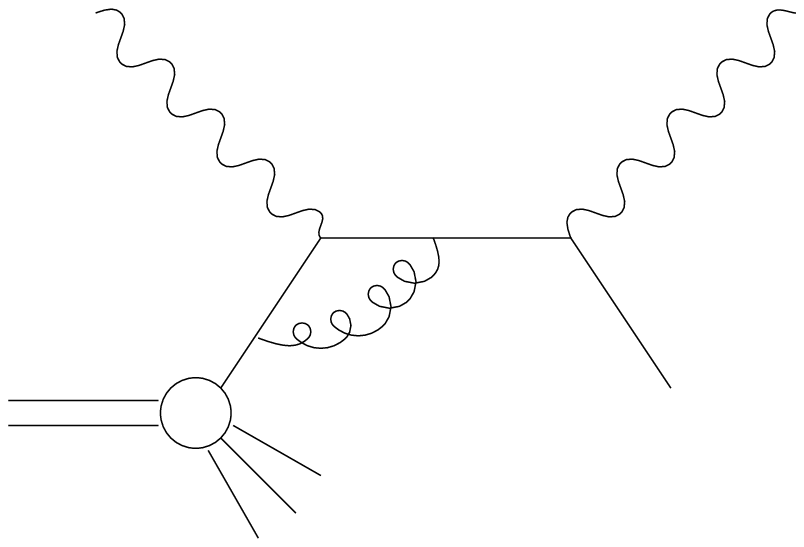}
\hskip -3cm
\epsfysize=3.5cm
\epsfbox{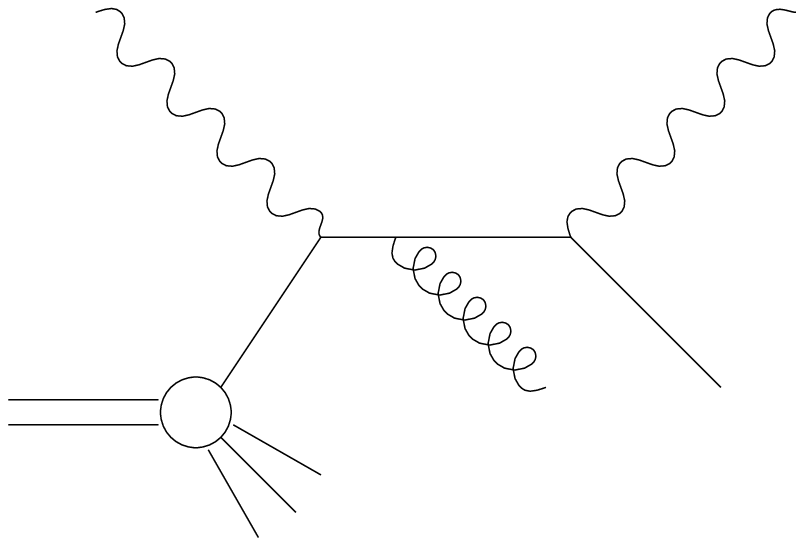}
\hskip 3.5cm
\epsfysize=3.cm
\epsfbox{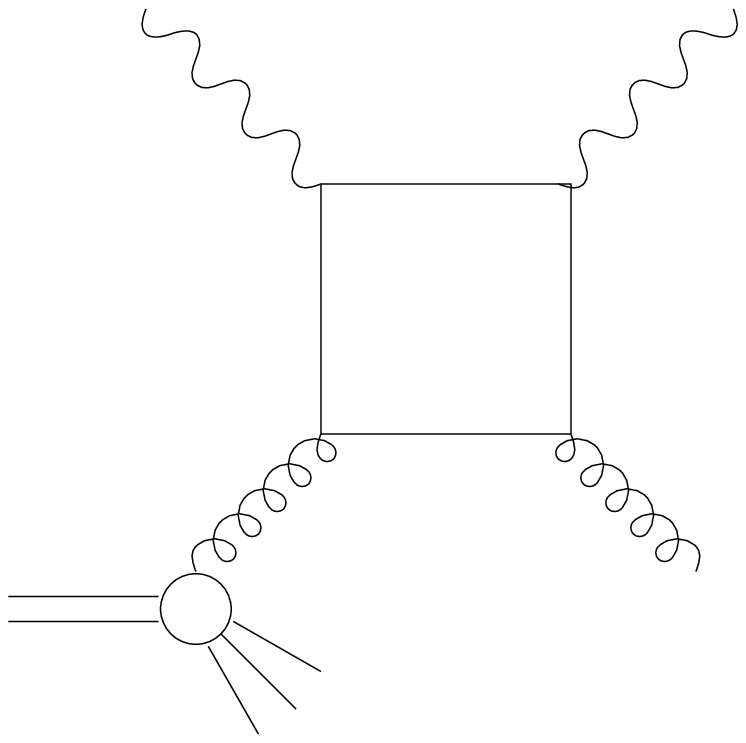}
\end{center}
\caption{Examples of the virtual gluon (left)  and real gluon (center) 
 $\alpha_s$ corrections to the Born contribution. Also a box diagram  
is shown (right).}
\label{fig:corbox}
\end{figure}

The Born level contribution to the cross section for the DIC process 
(\ref{eq:1}),
i.e. the lowest order in the strong coupling $\alpha_s$ term, arises from the 
Compton process on the quark (fig.~\ref{fig:dicborn},right ):
\be
\gamma q\rightarrow \gamma q.
\label{eq:2}
\ee
It gives the [$\alpha_{em}^2$] order contribution to the partonic cross 
section.
At the same $\alpha_{em}^2$ order it contributes to the hadronic 
cross section for the process $\gamma p\rightarrow \gamma X$.
The Parton Model (PM) prediction for the DIC process (\ref{eq:1}),
which applies for $x_T=2p_T/\sqrt{S}\sim{\cal O}(1)$,
relies solely on the Born contribution 
(\ref{eq:2})~\cite{Bjorken:1969ja}, namely:
\be
d\sigma^{\gamma p\rightarrow\gamma X}
=\sum_q\int dx_p q_p(x_p){d\hat {\sigma}^{\gamma q\rightarrow \gamma q}},
\label{eq:3}
\ee
where $q_p$ is the quark density in the proton and 
$d\sigma^{\gamma p\rightarrow\gamma X}$ 
($d\hat\sigma^{\gamma q\rightarrow \gamma q}$)
stands for the hadronic (partonic) cross section.
In the QCD improved PM the cross section is
given by (\ref{eq:3}), however with scale dependent quark densities. 
For semihard processes, where $x_T \ll 1$, the prediction
based on the process (\ref{eq:2}) only is not a sufficient approximation, 
and one
should also consider the contributions corresponding to the collinear showers, 
involving hadronic-like interactions of the photon(s). 
There are two classes of such contributions:
{\sl single resolved} photon processes with resolved initial $or$ 
final photon, and {\sl double resolved} photon processes with
both the initial $and$ the final photon resolved
(figs. \ref{fig:12res}).
They correspond to partonic cross sections of orders 
[$\alpha_{em} \alpha_s$] (single resolved)
and [$\alpha_s^2$] (double resolved).
If one takes into account that partonic densities in the photon and
the parton fragmentation into the photon are of order $\sim\alpha_{em}$, 
then the contributions to the hadronic cross section from these resolved 
photon processes are  $\alpha_{em}^2 \alpha_s$ and 
$\alpha_{em}^2 \alpha_s^2$, respectively. 
Both single and double resolved photon contributions are included in the 
standard LL QCD analyses of the DIC 
process~\cite{Duke:1982bj,Aurenche:1984hc,Gordon:1994sm}. 
To obtain the NLO QCD predictions for the process (\ref{eq:1})
the $\alpha_s$ corrections to the lowest order process (\ref{eq:2}) 
have to be calculated leading to terms of order 
$\alpha_{em}^2\alpha_s$~\cite{Duke:1982bj,Aurenche:1984hc,mkcorr,jan} 
(fig.~\ref{fig:corbox}).
In these $\alpha_{em}^2 \alpha_s$ contributions
there are collinear singularities to be subtracted and 
shifted into corresponding quark densities {\sl or} fragmentation functions.
This way the single resolved photon contribution appears
in the calculation of the $\alpha_s$ corrections to the Born process. 
It is worth noticing that in
the NLO expression for the cross section there are no 
collinear singularities which would lead to the double
resolved photon contributions.
It indicates that taking into account [$\alpha_s^2$] subprocesses,
associated with both the initial and final photons resolved,
goes beyond the accuracy of the NLO calculation.
This will be consistent
within the  NNLO approach, where $\alpha_s^2$ correction to the Born term and
$\alpha_s$ correction to the single resolved terms should be included,
all giving the same  $\alpha_{em}^2\alpha_s^2$ order contribution to the
hadronic cross sections.

The other set of diagrams is considered by some 
authors~\cite{Gordon:1995km}-\cite{Gordon:1998yt} 
and \cite{Fontannaz:2001ek,Fontannaz:2001nq} 
in the NLO approach to DIC process 
(\ref{eq:1}).
This approach, which we will call
``$1/\alpha_s$'' approach, is motivated by 
large logarithms of $Q^2$ in the $F_2^{\gamma}$ existing already in the 
PM.
By expressing  $\ln (Q^2/\Lambda_{QCD}^2)$ as $\sim{1/\alpha_s}$ 
one treats
the parton densities in photon (and parton fragmentation into the photon) as proportional to  $\alpha_{em}/\alpha_s$
(see e.g.~\cite{Fontannaz:1982et}-\cite{Aurenche:1984hc},\cite{Aurenche:1992sb,Gordon:1994sm},\cite{Gordon:1995km}-\cite{Gordon:1998yt} and
\cite{Fontannaz:2001ek,Fontannaz:2001nq} ).
By applying this method to the DIC process, we see that
the single resolved photon contribution to the  cross section
for the process $\gamma p\rightarrow\gamma X$
becomes of the same order as the Born term.The same is also observed for the 
double resolved photon contribution. Namely, we have for the Born, single and double resolved photon contributions:
$$
{1}\otimes [ \alpha_{em}^2] \otimes 1=\alpha_{em}^2, {\hspace{0.5cm}} {{\alpha_{em}}\over{\alpha_s}}\otimes [  \alpha_{em} \alpha_s]  
\otimes 1=\alpha_{em}^2, {\hspace{0.5cm}} 
{{\alpha_{em}}\over{\alpha_s}}\otimes [ \alpha_s^2]  \otimes 
 {{{\alpha_{em}}}\over{\alpha_s}}=
\alpha_{em}^2.
$$
In such counting,
 the same $\alpha_{em}^2$ order contributions
to the hadronic cross section are given by the direct Born process,
single and double resolved photon processes
although  they correspond to    quite different final states 
(observe a lack of the remnant of the photon in the direct process).
Moreover, they constitute the lowest order
(in the strong coupling constant) term in the perturbative expansion,
actually the zeroth order, so 
the direct dependence of the cross section
on the strong coupling constant is absent.
Some of these terms  correspond to 
the hard processes involving gluons, still there are no  terms proportional
to $\alpha_s$ coupling!

In the ``$1/\alpha_s$'' approach the $\alpha_s$ correction to
the Born cross section,  the single and to the 
double resolved photon contributions are included in the NLO calculation,
since all of them give terms of the same order,
$\alpha_{em}^2\alpha_s$, see ~\cite{Gordon:1995km}-\cite{Gordon:1998yt},
\cite{Fontannaz:2001ek,Fontannaz:2001nq}.

To summarize, the first approach starts with one basic, direct subprocess 
as in the PM (eq. \ref{eq:2}), while the second approach  with three different 
types of subprocesses (as in the standard LL calculation). 
Obviously, some of 
NNLO terms in the first method belong to the NLO terms in the second one.

In this paper we apply the first type of NLO approach to the DIC process,
in particular
%
we take
into account the following subprocesses:\\
$\bullet$ the Born contribution (\ref{eq:2}) (fig.~\ref{fig:dicborn}, left);\\
$\bullet$ the finite
$\alpha_s$ corrections to the Born diagram (so called K-term) from
virtual gluon exchange, real gluon emission (fig.~\ref{fig:corbox}, 
left and center), and 
the process $\gamma g \rightarrow q \bar q \gamma$; \\
$\bullet$ two types of single resolved photon contributions, 
with resolved initial and final photons
(fig.~\ref{fig:12res}, left and center).

Besides the above full NLO set, we will include two terms of order 
$\alpha_{em}^2 \alpha_s^2$ (formally  from the NNLO set): the double 
resolved contributions (fig.~\ref{fig:12res},right) and the 
direct diagram (box)
$\gamma g \rightarrow \gamma g$~\cite{Combridge:1980sx} 
(fig.~\ref{fig:corbox},right), 
since they were found to be 
large~\cite{Fontannaz:1982et}-\cite{Aurenche:1992sb}. 

The cross section for the
$\gamma p\rightarrow\gamma X$ scattering has the following form:
\bea
E_{\gamma}{d^3\sigma^{\gamma p\rightarrow\gamma X}\over
d^3p_{\gamma}} = \sum_{b}\int dx f_{b/p}(x,\bar{Q}^2) 
{\alpha_s(\bar Q^2)\over 2\pi^2 \hat{s}} K_b +  \nonumber \\
+\sum_{abc}\int {dz\over z^2}\int dx_{\gamma}\int dx 
f_{a/\gamma}(x_{\gamma},\bar{Q}^2) 
f_{b/p}(x,\bar{Q}^2)
\cdot
D_{\gamma /c}(z,\bar{Q}^2)
E_{\gamma}{d^3\sigma^{ab\rightarrow cd}\over d^3p_{\gamma}}.
\eea
The first term is the K-term, and the second one stands
 for the sum over all other
contributions including the Born term. 
The $f_{a/\gamma}$ ($f_{b/p}$) is a $a$ ($b$)-parton 
distribution in the photon (proton) while the $D_{\gamma /c}$ is a $c$-parton 
fragmentation function. For the direct initial (final) photon, 
where $a=\gamma$ ($c=\gamma$), we take $f_{a/\gamma}=\delta (x_{\gamma}-1)$
($D_{\gamma /c}=\delta (z-1)$) 
(the Born contribution is obtained for $a=\gamma$, $b=q$ and $c=\gamma$).
The variables $x_{\gamma}$, $x$ and $z$ 
stand for the fraction of the initial photon,
proton, and $c$-parton momenta taken by the $a$-parton, $b$-parton,
and the final photon, respectively.

\section{The isolation}
\label{sec:iso}
In order to observe photons originating from a hard subprocess
one should reduce backgrounds, mainly from $\pi^0$'s and $\gamma$'s
radiated from final state hadrons. To achieve this, isolation cuts on the
observed photon are introduced in experimental analyses. The isolation 
cuts are defined by demanding that the sum of transverse 
hadronic energy within a cone of radius $R$ around the final 
photon, where
the radius $R$ is defined in the rapidity and azimuthal
angle space,
should be smaller than the final photon transverse energy multiplied
by a small parameter $\epsilon$:
$
\sum_{hadrons}E_{Th}<\epsilon E_{T\gamma}$
\footnote{Some aspects of the QCD calculation of the isolated photon production
are discussed  in ~\cite{Berger:1996cc,Aurenche:1997ng}.}.

The simplest way to calculate the differential cross section
for an isolated photon, $d\sigma_{isol}$, is to calculate
the difference of a non-isolated  differential cross section,
$d\sigma_{non-isol}$, and a subtraction term, which corresponds to cuts 
opposite to the isolation cuts
$d\sigma_{sub}$~\cite{Berger:1990es}-\cite{Gluck:1994iz,Gordon:1995km}:
$
d\sigma_{isol}=d\sigma_{non-isol}-d\sigma_{sub}.
$


Note that in practice the isolation cuts are imposed only when calculating the K-term, and 
the  contributions involving fragmentation function (resolved final photon). 
In calculation of the subtraction term for the $K$-term we applied  
a small-$\delta$ approximation, 
see~\cite{Gordon:1994ut,Gordon:1995km} ~\footnote{
This small $\delta$ approximation seems to be  an accurate
analytic technique for including isolation effects in NLO calculations
(also for $R$ = 1), see discussion in ~\cite{Gordon:1998yt}.}.
 Presently we are working on 
calculations of the cross sections for the $\gamma$ and $\gamma$+jet
photoproduction  using for a comparison the space slicing method ~\cite{z}, 
as in ~\cite{Gordon:1998yt,Fontannaz:2001ek,Fontannaz:2001nq}.
\section{The results and comparison with data}\label{results}

We consider the production of photons with large transverse momentum,
$p_T\gg\Lambda_{QCD}$, in the $ep$ scattering, $ep\rightarrow e\gamma X$,
at the HERA collider using the equivalent photon
(Williams-Weizs\"{a}cker) approximation~\cite{vonWeizsacker:1934sx}: 
\bea
d{\sigma}^{ep\rightarrow e\gamma X}=\int G_{\gamma/e}(y) 
d{\sigma}^{\gamma p\rightarrow\gamma X} dy ,
\eea
where $y$ is (in the laboratory frame) a fraction of the initial electron 
energy taken by the photon~\cite{Budnev:1974de}:
\bea
G_{\gamma/e}(y)={\alpha_{em}\over 2\pi} \{ {1+(1-y)^2\over y}
\ln [ {Q^2_{max}(1-y)\over m_e^2 y^2}] 
- ~ {2\over y}(1-y-{m_e^2y^2\over Q^2_{max}}) \},
\eea
with $m_e$ being the electron mass. 
We assume $Q^2_{max}$ equal to  1 GeV$^2$, as  in the 
recent photoproduction measurements at the HERA collider.
We neglect the large $p_T$ photon emission from the 
electron~\cite{ula}.

The results for the non-isolated and isolated photon cross sections are 
obtained in NLO accuracy with additional NNLO
terms, as discussed above. 
We take the HERA collider energies: $E_e$=27.5 GeV and 
$E_p$=820 GeV~\cite{Breitweg:2000su},
and  the $p_T$ range of the final photon between 
5 and 20 GeV ($x_T$ from 0.03 to 0.13).
The ${ \overline {\rm MS}}$ scheme with
a hard (renormalization, factorization) scale $\bar{Q}$ equal $p_T$
(also $\bar{Q}=p_T/2$ and $2p_T$)
was applied.
We assume
the number of active (massless) flavors 
to be $N_f$=4 (and for comparison also $N_f$=3 and 5).
The two-loop coupling constant $\alpha_s$ is used
with $\Lambda_{QCD}$=0.365, 0.320 and 0.220 GeV for $N_f$=3, 4 and 5,
respectively, as fitted by us 
to the experimental value of $\alpha_s(M_Z) = 0.1177$~\cite{Biebel:1999zt}.

We use the GRV parametrizations of the proton
structure function (NLO and LO)~\cite{Gluck:1995uf}, 
the photon structure function (NLO and LO)~\cite{Gluck:1992ee},
and the fragmentation function (NLO)~\cite{Gluck:1993zx}. For comparison
other parametrizations were also 
used: DO~\cite{Duke:1982bj}, ACFGP~\cite{Aurenche:1992sb}, 
CTEQ~\cite{Lai:1997mg}, MRST~\cite{Martin:1998sq} and GS~\cite{Gordon:1997pm}.
As the reference we take the GRV NLO set of parton 
distributions~\cite{Gluck:1995uf}-\cite{Gluck:1993zx},
$N_f = 4$, $\Lambda_{QCD} = 320$ GeV and $\bar{Q}=p_T$.

\subsection{Non-isolated versus isolated photon cross section}
\label{results1}

We have studied the $p_T$ distribution for the produced final photon 
without any cut
and found that it  decreases by three orders of magnitude when $p_T$
increases from 4 GeV to 20 GeV (not shown). Obviously the most important 
contribution is coming from the lowest $p_T$ region, 
where the resolved photon processes dominate.
The total NLO  cross section integrated over $p_T$ range from 5 to 10 GeV, 
is equal to 226 pb, with  
individual contributions  equal to:
$Born=36.3\% $, $single$ $resolved=35.1\% $,
$double$ $resolved=18.7\%$, $box=6.2\% $, $K$-$term$=3.9\%,
so the single resolved photon processes
give a contribution comparable to the Born term. Also the double resolved
 photon processes are  important. 
The direct box diagram ($\gamma g\rightarrow\gamma g$) 
gives 17\% of the Born ($\gamma q\rightarrow\gamma q$) contribution. Such
relatively large  contribution 
is partially due to large gluonic content of the proton at small $x_p$.

The differential cross section for the final photon rapidity, 
$d\sigma /d\eta_{\gamma}$, for the non-isolated photon and  
 for the  photon isolated with various  
 cones (various $\epsilon$, $R$) was studied. 
The isolation cut suppresses the cross section significantly in the whole 
rapidity range(not shown). For $\epsilon$=0.1 and $R=1$
the suppression is 17-23\% at rapidities $-1.5<\eta_{\gamma}\le 4$
\footnote{The positive
rapidity is pointed in the proton direction.}.


As expected, the cross section for fragmentation processes 
 is strongly suppressed: 
after isolation it is lowered by a factor of 5. 
At the same time the QCD corrections to the Born diagram increase
significantly, i.e. the contribution to the subtraction
cross section, $d\sigma_{sub}$, due to this corrections is negative.
The subtraction cross section, being a sum of negative QCD corrections
and fragmentation contributions, is of course positive.

\subsection{Other experimental cuts}\label{results1b}

In order to compare the results with data we fix  $R$=1 and 
$\epsilon$=0.1,
which are the standard values used in both theoretical and experimental 
analyses, and consider other 
cuts imposed by the ZEUS group ~\cite{Breitweg:2000su}.
Two types of the final state were measured in the ZEUS experiment:
1) an isolated photon with $-0.7\le\eta^{\gamma}\le 0.9$ 
and $5\le p_T\le 10$ GeV; 
2) an isolated photon plus jet with the photon 
rapidity and transverse momentum as above, the jet rapidity in the range
$-1.5\le\eta^{jet}\le 1.8$, and the jet transverse momentum 
$p_T^{jet}\ge 5$ GeV.
Here we compare our NLO predictions with the ZEUS 
data from the first type of measurements~\cite{Breitweg:2000su}.
More results can be found in \cite{zk} and \cite{z}.

We have found (see \cite{zk}) that the cross section for a 
production of final $\gamma$ is strongly reduced, by 30-85\%, 
in the positive rapidity region due to   the limited energy range, 
$0.2\le y\le 0.9$ (see also 
~\cite{Fontannaz:2001ek,Fontannaz:2001nq} for a similar conclusion).
 At negative rapidities the change due to the $y$-cut is weaker: 5-10\% 
at $-1.2<\eta_{\gamma}\le -0.4$ and 10-30\% at other negative rapidities.
Note however, that the Born term  is reduced 3.5 times. 

\begin{figure}
\hskip 1cm
\begin{center}
\epsfysize=6cm
\epsfbox{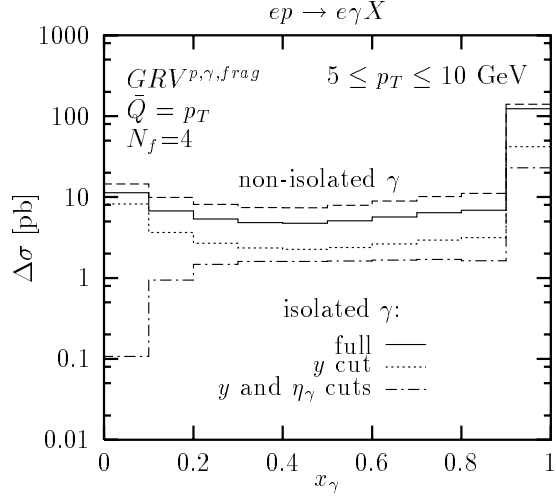}
\end{center}
\caption{\label{fig:dxp}The cross section in $x_{\gamma}$ bins of the 
length 0.1. 
The results for non-isolated $\gamma$ 
are shown (dashed line). 
The solid line represents results for isolated $\gamma$ 
with $\epsilon = 0.1$ and $R = 1$.
Results with additional cuts in the
isolated $\gamma$ cross section are shown with: dotted line
($0.2\leq y\leq 0.9$) and dot-dashed line ($0.2\leq y\leq 0.9$,
$-0.7\leq\eta_{\gamma}\leq 0.9$).}
\end{figure}

The role of various experimental cuts is illustrated
 in fig.~\ref{fig:dxp}, for the $x_{\gamma}$ distribution.
The small $x_{\gamma}$
contributions are strongly, by two orders of magnitude, 
diminished by the photon rapidity cut.
This shows that measurements at the central $\eta^{\gamma}$ region 
($-0.7\le\eta^{\gamma}\le 0.9$)
are not too sensitive to the small $x_{\gamma}$ values in the photon,
and can not be  used presently for constraining eg. 
the gluon density in the photon.
\subsection{The comparison with data}\label{results2}

\begin{figure}
\begin{center}
\epsfysize=5cm
\epsfbox{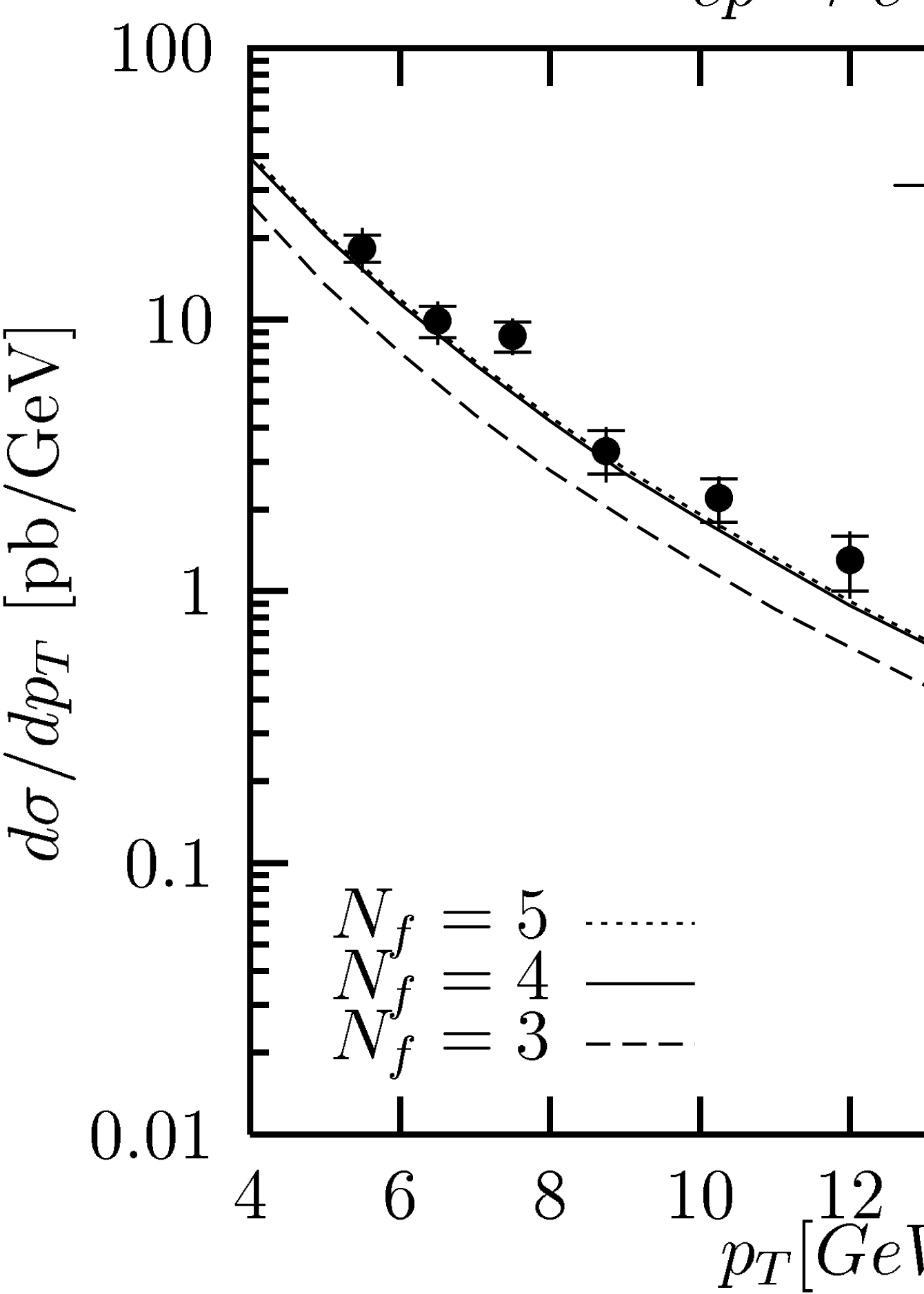}
\hskip -3.5cm
\epsfysize=5cm
\epsfbox{kraw.fig5r.ps}
\hspace*{6cm}
\end{center}
\caption{\label{fig:dnNf}
The results for isolated $\gamma$ for:
$N_f$ = 3 (dashed lines), 4 (solid lines) and 5 (dotted lines),
compared to the ZEUS data~\protect\cite{Breitweg:2000su}.
 The  $d\sigma/dp_T$ as a function of the
photon transverse momentum (left) and
 $d\sigma/d\eta_{\gamma}$
as a function of the photon rapidity $\eta_{\gamma}$ (right);
the result without the box contribution
is also shown for $N_f$ = 4 (dot-dashed).} 
\end{figure}

In fig.~\ref{fig:dnNf} (left) the comparison is made with ZEUS data for
the $p_T$ distribution for isolated $\gamma$ for 
various $N_f$. A satisfactory 
agreement is obtained for $N_f=4$ (and 5).
Note  large difference between the results for $N_f$=4 and 3
due to the fourth power of electric charge.
The rapidity 
distribution is shown in fig.~\ref{fig:dnNf}(right), where  a good description 
of the data is obtained for $N_f$=4 and 5 in the rapidity region 
$0.1\le\eta_{\gamma}\le 0.9$. For $-0.7\le\eta_{\gamma}\le 0.1$ our 
predictions lie mostly below the experimental points. 
This disagreement between predicted and measured cross sections is observed 
also for other theoretical calculations (LG,FGH) and for Monte Carlo 
simulations~\cite{Breitweg:2000su}. 
In fig.~\ref{fig:dnNf}(right) we present separately an effect due to the box 
subprocess (for $N_f=4$). It is clear that the box term enhances considerably 
the cross section in the measured rapidity region (by $\sim 10\%$).
 The double resolved photon 
contribution is also sizeable, although roughly two times smaller than the 
box one.
Both these $[\alpha_s^2]$ contributions improve description of the data.

\begin{figure}
\begin{center}
\hspace*{0.5cm}
\epsfysize=5cm
\epsfbox{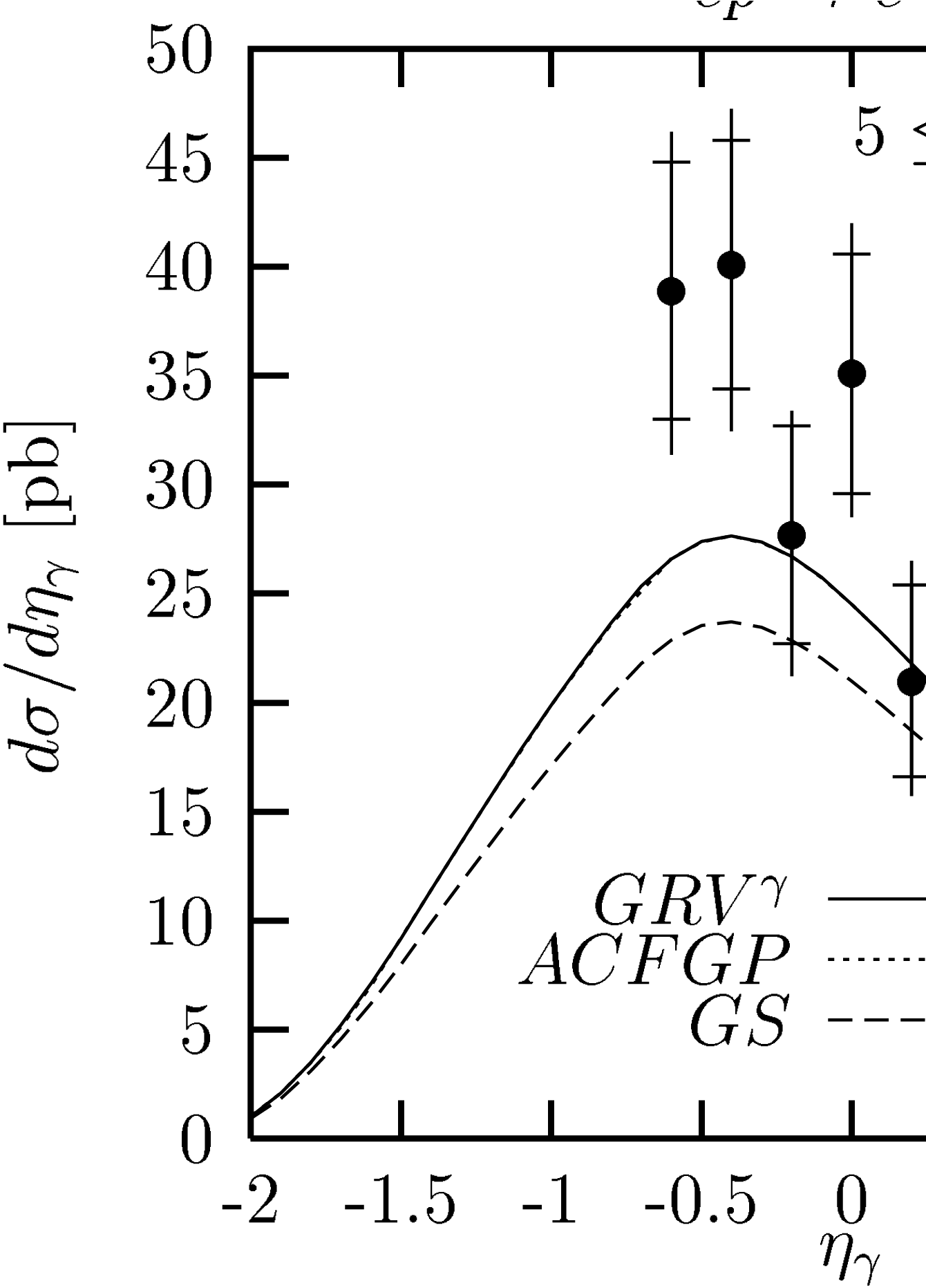}
\hskip -3.5cm
\epsfysize=5cm
\epsfbox{kraw.fig6r.ps}
\hspace*{6cm}
\end{center}
\caption{\label{fig:dnpar}The  $d\sigma/d\eta_{\gamma}$
for isolated $\gamma$ compared to
the ZEUS data~\protect\cite{Breitweg:2000su}.
Three different NLO  parton distributions for $\gamma$: 
ACFGP~\protect\cite{Aurenche:1992sb} 
(dotted line), GRV~\protect\cite{Gluck:1992ee} (solid line) and 
GS~\protect\cite{Gordon:1997pm} (dashed line).
 $\bar{Q} = p_T$ (left) and $\bar{Q} = 2 p_T$ (right). }
\end{figure}

\begin{figure}
\begin{center}
\epsfysize=5.5cm
\epsfbox{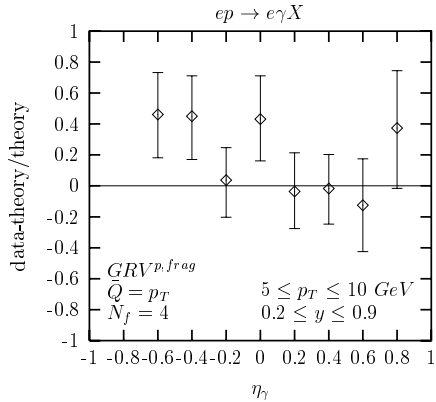}
\end{center}
\caption{\label{fig:exth} (Data-theory)/theory is plotted for GRV 
parton parametrizations for the initial and final photons, and for proton, the 
ZEUS data from ~\protect\cite{Breitweg:2000su}.}
\end{figure}

\begin{figure}
\begin{center}
\hskip -7.cm
\epsfysize=5cm
\epsfbox{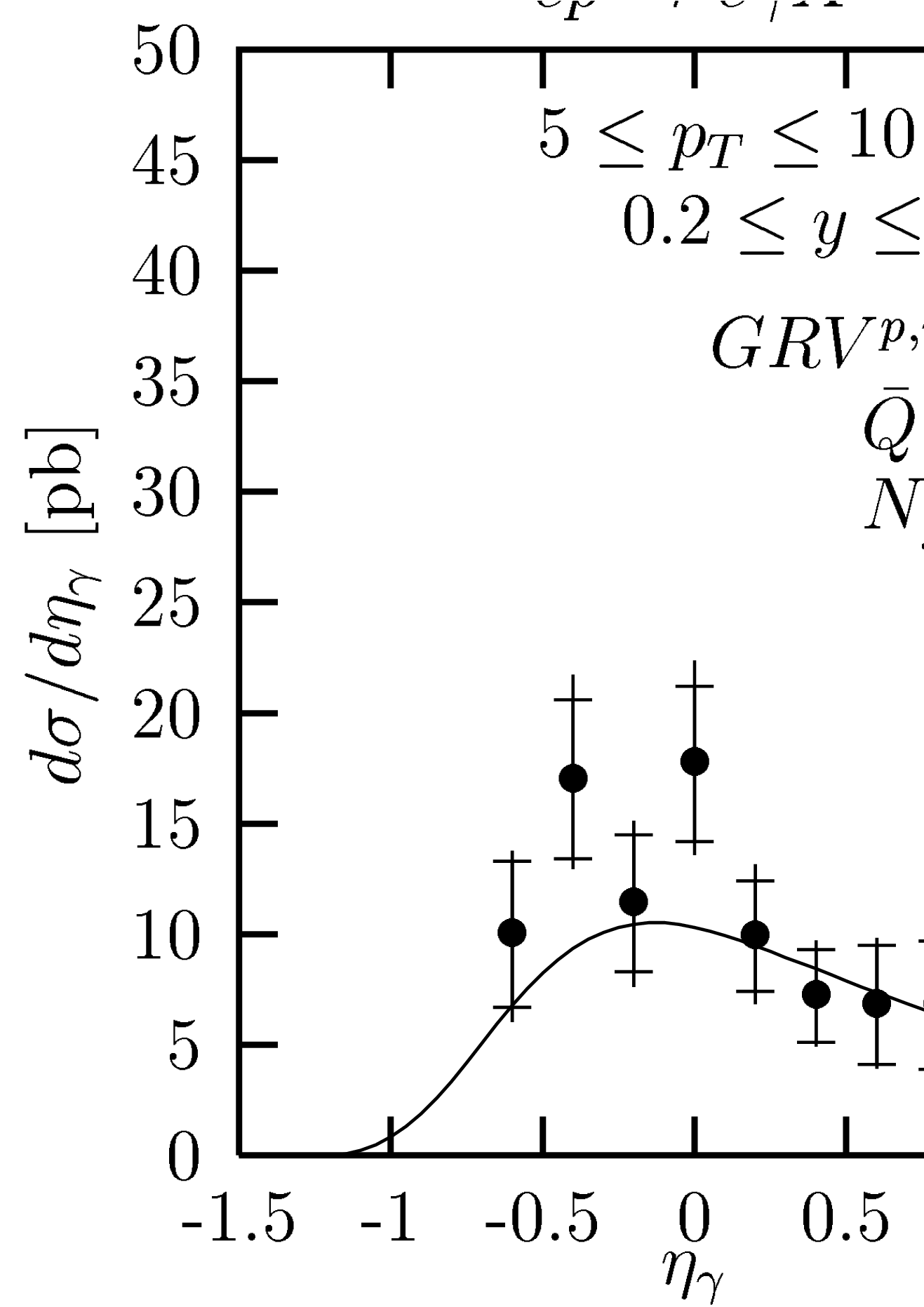}
\hskip -3.5cm
\epsfysize=5cm
\epsfbox{kraw.fig8c.ps}
\hskip -3.5cm
\epsfysize=5cm
\epsfbox{kraw.fig8r.ps}
\end{center}
\caption{The results for three ranges of $y$:
$0.2<y<0.32$, $0.32<y<0.5$ and $0.5<y<0.9$, compared to the ZEUS 
data~\protect\cite{Breitweg:2000su}.}
\label{fig:bins}
\end{figure}

The predictions obtained using three different NLO parton densities in the 
photon (ACFGP~\cite{Aurenche:1992sb}, GRV~\cite{Gluck:1992ee} and 
GS~\cite{Gordon:1997pm}) are presented for $N_f=4$
in fig.~\ref{fig:dnpar} for $\bar{Q} = p_T$ (left) and for 
$\bar{Q} = 2 p_T$ (right) together with the ZEUS data~\cite{Breitweg:2000su}.
The results based on ACFGP and GRV parametrizations differ by less than 4\%
at rapidities $\eta_{\gamma} < 1$ (at higher $\eta_{\gamma}$ the difference is 
bigger), and both give good description of the data in the rapidity range
$0.1\le\eta_{\gamma}\le 0.9$ (for $\bar{Q} = p_T$ and $\bar{Q} = 2 p_T$). 
For $-0.7\le\eta_{\gamma}\le 0.1$ none of the predictions is in
agreement with the measured cross section.
For $\bar{Q}$=$p_T$ (fig.~\ref{fig:dnpar}left) the GS distribution leads to 
results considerably below ones obtained using ACFGP and GRV densities,
especially in the rapidity region from roughly -1 to~1.
This 
is  due to a different treatment of the 
charm quark in the photon,
namely in the GS approach the charm quark is absent
for $\bar{Q}^2$ below 50 GeV$^2$
- contrary
to GRV and ACFGP parametrizations  where the charm threshold occurs at lower $\bar{Q}^2$.
All the considered parton distributions give similar description of the data
when the scale is changed to $\bar{Q}=2p_T$,
since then $\bar{Q}^2$ is always above 50 GeV$^2$.
The (data-theory)/theory for the same cross section for the reference set (GRV)
 of parton parametrizations is presented in 
 fig.~\ref{fig:exth}.
In fig.~\ref{fig:bins} our predictions are compared to the ZEUS data
divided into three ranges of $y$. Clearly  
the discussed above discrepancy between the data and the predictions for 
$\eta^{\gamma} < 0.1$ is coming mainly from the low $y$ region, $0.2<y<0.32$.
In the high $y$ region, $0.5<y<0.9$, a good agreement is obtained
for a whole measured rapidity region.
This is not the case of LG and FGH results, which are  in disagreement
even for a large $y$ (mainly for  rapidities above 0.1,  see below).

We have also studied the dependence of our results on the choice of 
the parton distributions in the proton and parton fragmentation
into the photon (not shown), and a small sensitivity was found.
Only at minimal ($\eta_{\gamma}<-1$)
and maximal ($4<\eta_{\gamma}$) rapidity values
this difference is larger, being at a level of $3.5-8\%$.
\section{The theoretical uncertainties  and comparison 
with other NLO analyses}\label{th}



In order to estimate the contribution due to missing higher order 
terms, we have studied the influence of the choice of the $\bar Q$ scale
 for the $\eta_{\gamma}$ distribution.  Some of results can be found 
in fig.~\ref{fig:dnpar} for the GRV and ACFGP parton  parametrizations.
Around the maximum of the cross section at rapidities 
$-1\le\eta_{\gamma}\le 0$ changing $\bar Q$ scale from $p_T/2$ to $2p_T$
leads to differences  4-6\%. 
This small sensitivity of the results to the change of the scale is 
important since it indicates that the contribution from neglected NNLO
and higher order terms is not significant. 
Note that individual contributions are strongly dependent
on the choice of $\bar{Q}$, e.g. results for the
single resolved processes vary by
$\pm$10-20\% at rapidities $\eta_{\gamma}\le 1$.
Results are much more stable
only when the sum of resolved processes
and QCD corrections is considered.
These results leads to expectation that our prediction
 should not differ too much from results based on larger set of diagrams.

As we discussed in Sec.~\ref{sec:dic}, our NLO calculation of the DIC process 
differs from the ``$1/\alpha_s$''-type NLO analysis
presented in ref.~\cite{Gordon:1995km}-\cite{Gordon:1998yt} and
 ~\cite{Fontannaz:2001ek,Fontannaz:2001nq},
by set of diagrams included in the calculation.
We do not take into account $\alpha_s$ corrections to the single and double
resolved processes, which are beyond the NLO accuracy in our approach. 
On the other hand, we include the box diagram neglected
in~\cite{Gordon:1995km}-\cite{Gordon:1998yt}, which however is taken into account in FGH analysis  ~\cite{Fontannaz:2001ek,Fontannaz:2001nq}. 
(The double resolved subprocesses are 
included in all mentioned   analyses.) 
\begin{figure}
\begin{center}
\hspace*{1.5cm}
\epsfysize=6cm
\epsfbox{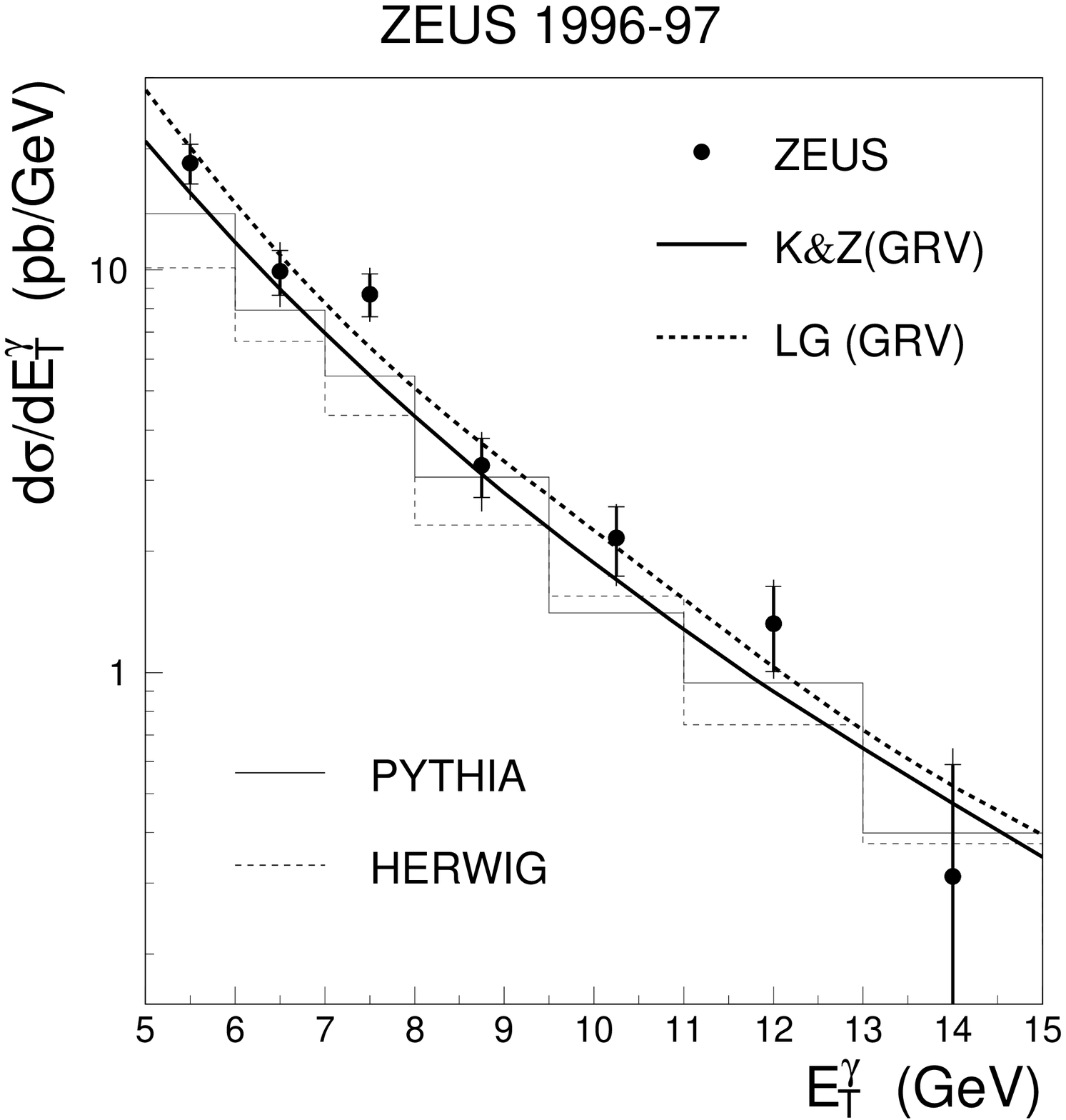}
\hskip 1cm
\epsfysize=6cm
\epsfbox{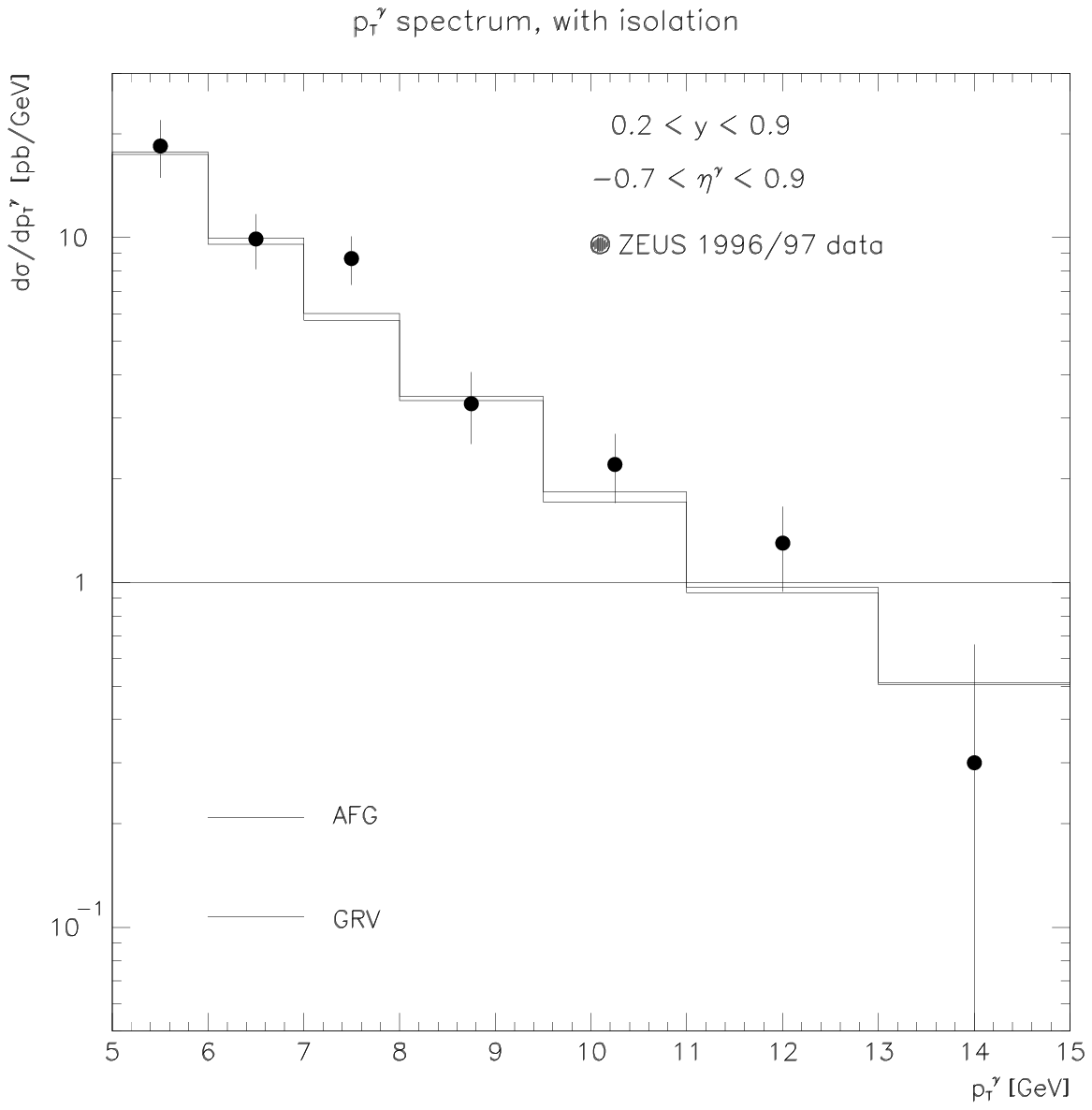}
\end{center}
\caption{ \label{fig:ept} The comparison of the data for the $p_T$ distribution (from ~\protect\cite{Breitweg:2000su})  with  PYTHIA and HERWIG Monte Carlo 
 predictions and  with our results and LG ones based on GRV parametrizations (left). The same data in comparison with the FGH results (left, from  ~\protect\cite{Fontannaz:2001ek}).} 
\end{figure}

\begin{figure}
\begin{center}
\hspace*{1cm}
\epsfysize=7cm
\epsfbox{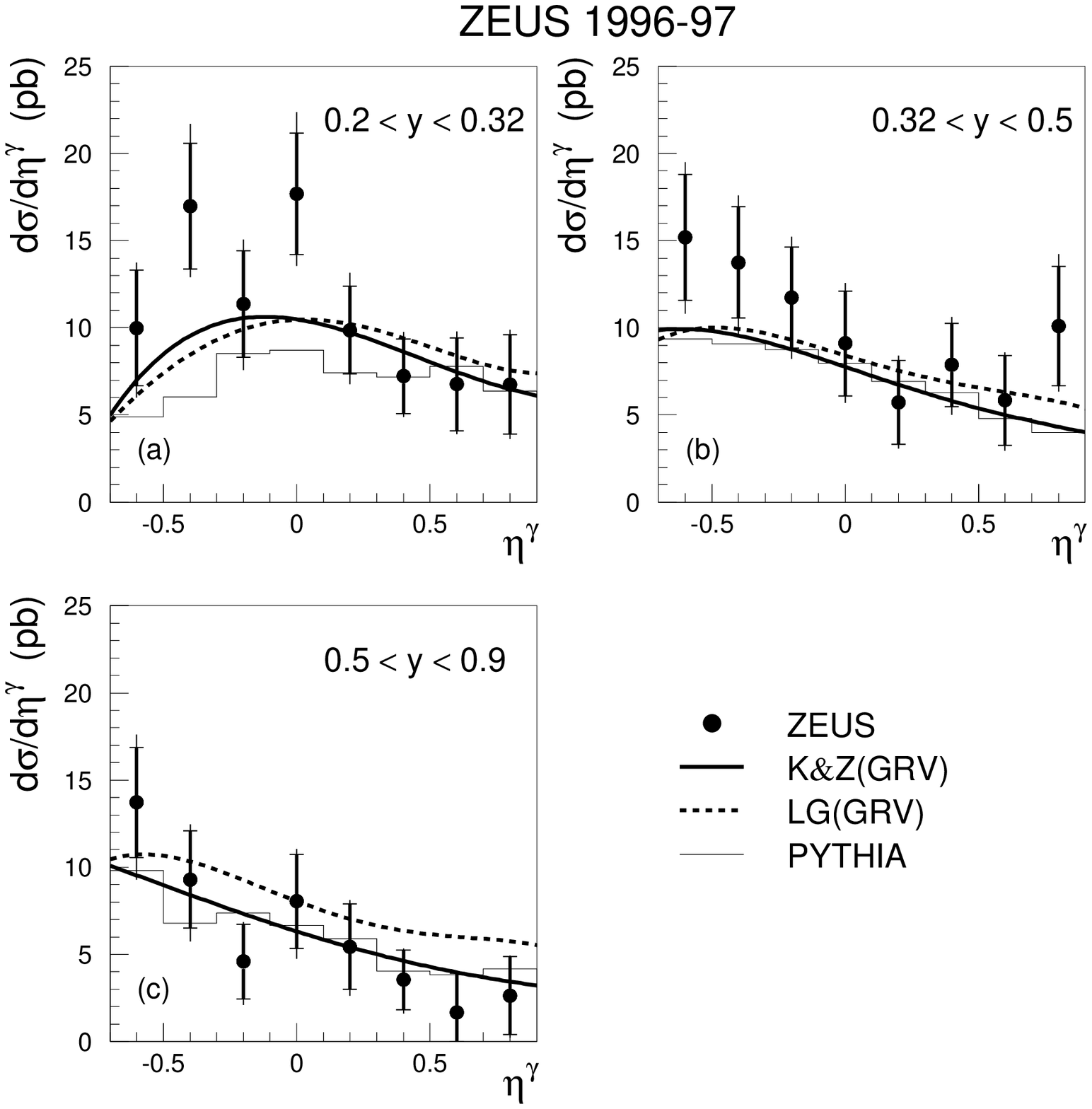}
\epsfysize=7cm
\epsfbox{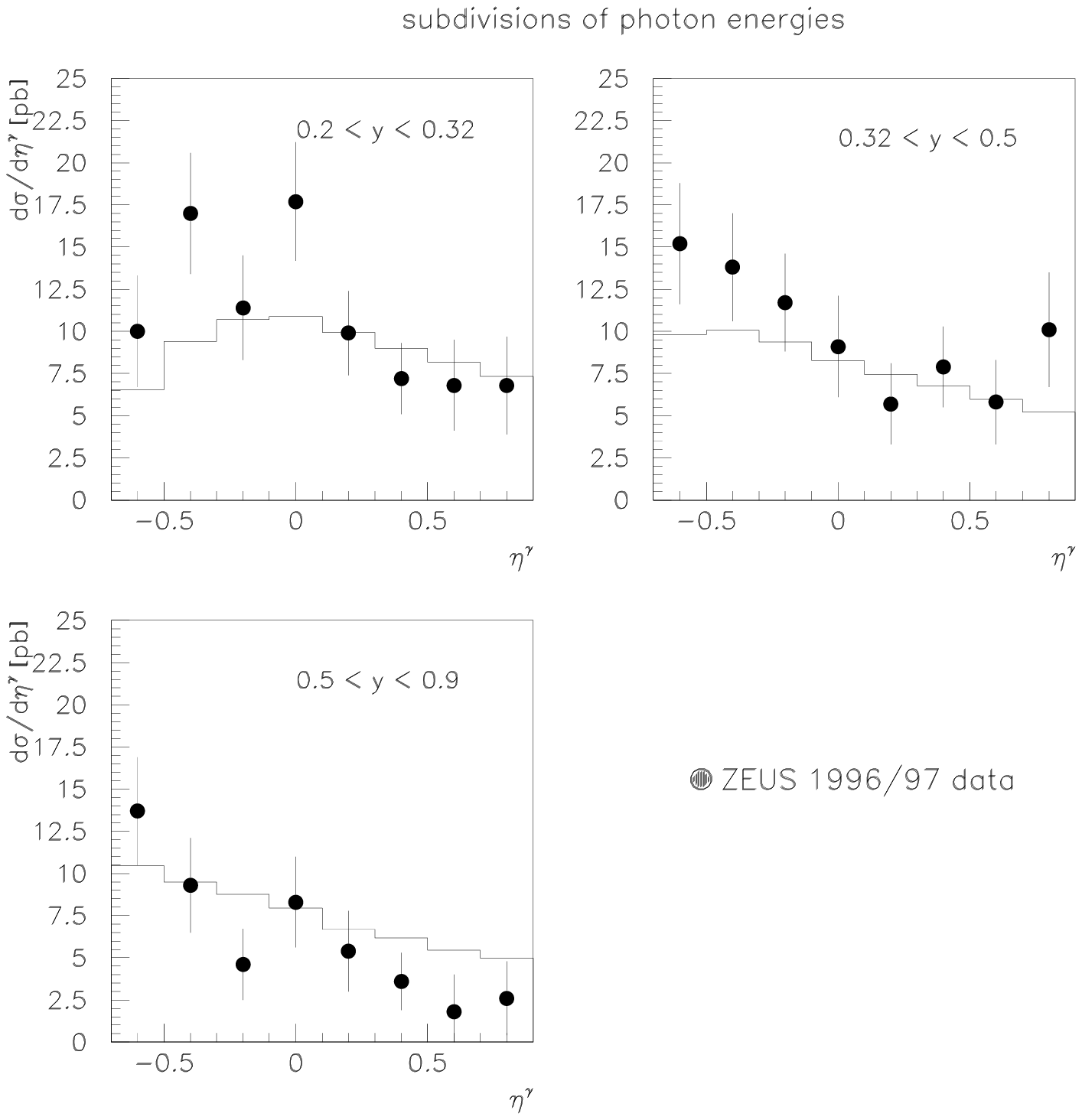}
\end{center}
\caption{\label{fig:eta3} The comparison with data for the rapidity distribution of the final photon three ranges of $y$ 
(as in fig.7 ) of PHYTIA prediction, our results and LG results (for GRV), 
 (left,  (from ~\protect\cite{Breitweg:2000su}))
 and similar comparison between data and FGH results, from  
~\protect\cite{Fontannaz:2001ek} (right).}
\end{figure}

We compare now our results and the results of the LG 
calculation~\cite{Gordon:1998yt} (using $N_f$=4 and $\bar{Q}=p_T$)
for the isolated final photon ($R$=1, $\epsilon$=0.1) in the 
kinematical range as in the ZEUS analysis~\cite{Breitweg:2000su}  
(i.e. for $-0.7\le\eta^{\gamma}\le 0.9$ and $0.2\le y\le 0.9$), see 
fig.~\ref{fig:ept}(left). In fig.~\ref{fig:ept}(right) a  comparison is made
for the FGH results ~\cite{Fontannaz:2001ek} and the same data. 
The LG predictions (and  FGH results) for $d\sigma /dp_T$ 
(with GRV parton parametrization) are about 20\% higher 
than ours in the presented range of transverse momentum, $4\le p_T\le 20$ GeV.
For $d\sigma /d\eta^{\gamma}$ cross section (with $5\le p_T\le 10$ GeV)
the biggest differences between our and LG  predictions are at large $y$ range,
what can be seen in fig.~\ref{fig:eta3} (left).
For $y$ range limited to low values only, 0.2 $< y <$ 0.32, the LG cross 
section is higher than ours by up to 20\% at positive $\eta_{\gamma}$, 
while at negative $\eta_{\gamma}$ it is lower by up to 10\%.
For large $y$ values, 0.5 $< y <$ 0.9,
where our predictions agree with data, the LG results are higher than 
ours by up to 80\% (at $\eta_{\gamma}$ = 0.9).
Not only the LG predictions  are too high as compared to the data in the 
forward direction, similar effect especially for large $y$
is seen in the  fig.~\ref{fig:eta3}(right), 
where the FGH results are compared with data.
Our predictions as we discussed above are close to the data for $\eta_{\gamma}$
above 0.1, see figs.~\ref{fig:exth},~\ref{fig:bins} and \ref{fig:eta3}(left).
\section{Summary}\label{sec:sum} 
Results of the NLO calculation, with additional NNLO contributions from double 
resolved photon processes and box diagram, for the isolated $ \gamma$ 
production in the DIC process at HERA are presented~\footnote{Our
fortran code is available upon request from azem@fuw.edu.pl.}.
The role of the 
kinematical cuts used in the ZEUS measurement~\cite{Breitweg:2000su} 
are studied in detail.

The results obtained using GRV parametrizations agree with the data
in shape and normalization for $p_T$ distribution.
For $\eta^{\gamma}$ distribution a good description of the data
is obtained for $\eta^{\gamma}>0.1$, while for $\eta^{\gamma}<0.1$
the data usually lie above the predictions. 
This discrepancy arises mainly from the low $y$ region, $0.2\le y\le 0.32$.
The beyond NLO terms,
especially a box contribution, improve the description of the data. 

We have studied the theoretical uncertainty of results due to the choice of 
the renormalization/factorization scale: $\bar{Q} = p_T/2, p_T, 2p_T$.
At high rapidities $\eta_{\gamma}> 3$, where the cross section
is small, this uncertainty is 10-30\%.
In a wide range of rapidities, $-2\le\eta_{\gamma}\le 2$, the dependence
on the $\bar{Q}$ scale is small, below 6\%.
Since we include some NNLO diagrams in our NLO calculation,
this stability of the predictions versus the change of the scale is 
especially important. The week dependence on the $\bar{Q}$ scale,
and not large differences between LL and NLO predictions (below 20\%)
allows to conclude that theoretical uncertainties of our NLO calculations
for an isolated photon production in the DIC process at HERA 
are relatively small.

We compared our results with the ``$1/\alpha_s$'' NLO calculations 
by LG and FGH, which are 
 based on different set of subprocesses.
The cross section $d\sigma/dp_T$ obtained by LG is about 20\% higher than 
ours (for GRV photonic parton distributions), FGH prediction is  closer to ours
than the LG one, it lies between KZ and LG curves.  For the cross section
$d\sigma/d\eta_{\gamma}$ the difference between our results and LG/FGH ones
 is up to 35\% at $\eta_{\gamma} = 0.9$.
The highest differences are present for high $y$ values only,
$0.5 < y < 0.9$, where on the other hand our predictions are in
agreement with the data. At low $y$ range, $0.2 < y < 0.32$, differences
between our calculation and calculations done by LG and FGH
 are smaller and none of them
describe the data well for rapidities below 0.1.
\section{Acknowledgments}
I would like to thank Andrzej Zembrzuski for his help in 
preparation of this talk and valuable discussions. 
The discussion with Jiri Chyla, Andreas Vogt,
Bernd Kniehl,  Michel  Fontannaz and Werner Vogelsang is  acknowledged.
I would like to thank organizers for this stimulating workshop.\\
Supported in part by Polish State Committee for Scientific
Research, grants number  2P03B05119 (2000-2001), 
  and by European Commission 50th framework
contract HPRN-CT-2000-00149.

\end{document}